\documentclass[preprint,5p,twocolumn,10pt]{elsarticle}

\usepackage[pdftex,hypertexnames=false,linktocpage=true]{hyperref}
\hypersetup{colorlinks=true,linkcolor=red,anchorcolor=darkgreen,citecolor=green!50!blue,filecolor=darkgreen,urlcolor=green,bookmarksnumbered=true,pdfview=FitB}

\usepackage{lipsum}
\usepackage{mathtools,cuted}

\usepackage{amsmath}

\usepackage{graphicx}

\usepackage{array}
\usepackage{booktabs}
\usepackage{arydshln}
\usepackage{multirow}

\newcommand{\ie}{\textit{i.e.}}

\usepackage{amsmath,amssymb,mathtools} 
\ifpdftex
    \usepackage{bbm}

\else
    \usepackage[bold-style=ISO]{unicode-math}
\fi

\usepackage{siunitx}  
\makeatletter
\newcommand{\vast}[1]{\bBigg@{#1}}
\makeatother

\newcommand{\intd}[3][]{\ifthenelse{\isempty{#3}}{\mathrm{d}^{#1} #2}{\frac{\mathrm{d}^{#1} #2}{#3}}\;}

\usepackage{stackengine}

\newcommand{\cond}{\; | \;}
\newcommand{\setcond}[2]{\ifthenelse{\isempty{#2}}{\{#1\}}{\{#1\cond{}#2\}}}

\usepackage{braket}
\usepackage{cancel}
    
\usepackage{braket}

\graphicspath{{../../plots/}}
\allowdisplaybreaks

\begin{document}

\begin{frontmatter}
\title{Transverse structure of the proton beyond leading twist: A light-front Hamiltonian approach}

\author[imp,ucas,keylab]{Zhimin Zhu}
\ead{zhuzhimin@impcas.ac.cn}

\author[imp,ucas,keylab]{Siqi Xu}
\ead{xsq234@impcas.ac.cn}

\author[imp,ucas,keylab]{Jiatong Wu}
\ead{wujt@impcas.ac.cn}

\author[ynu]{Hongyao Yu}
\ead{yuhongyao@stu.ynu.edu.cn}

\author[kek,imp]{Zhi Hu\corref{cor1}}
\ead{huzhi0826@gmail.com}

\author[imp,ucas,keylab,ashu]{Jiangshan Lan}
\ead{jiangshanlan@impcas.ac.cn}

\author[imp,ucas,keylab]{Chandan Mondal}
\ead{mondal@impcas.ac.cn}

\author[imp,ucas,keylab]{Xingbo Zhao}
\ead{xbzhao@impcas.ac.cn}

\author[iowa]{James P. Vary}
\ead{jvary@iastate.edu}

\author[]{\\\vspace{0.2cm}(BLFQ Collaboration)}

\address[imp]{Institute of Modern Physics, Chinese Academy of Sciences, Lanzhou, Gansu, 730000, China}
\address[ucas]{School of Nuclear Physics, University of Chinese Academy of Sciences, Beijing, 100049, China}
\address[keylab]{CAS Key Laboratory of High Precision Nuclear Spectroscopy, Institute of Modern Physics, Chinese Academy of Sciences, Lanzhou 730000, China}
\address[ynu]{Department of Physics, Yunnan University, Kunming, Yunnan 650091, China}
\address[kek]{High Energy Accelerator Research Organization (KEK), Ibaraki 305-0801, Japan}
\address[ashu]{Affiliated School of Huizhou University, Huizhou, Guangdong, 516001, China}
\address[iowa]{Department of Physics and Astronomy, Iowa State University, Ames, IA 50011, USA}

\cortext[cor1]{Corresponding author}

\begin{abstract}
  Within the Basis Light-Front Quantization framework, we systematically investigate the subleading twist (twist-3) transverse-momentum-dependent parton distribution functions (TMDs) of the proton beyond the Wandzura-Wilczek (WW) approximation. 
  The subleading twist TMDs are not independent and can be decomposed into twist-2 and genuine twist-3 terms from the equations of motion. 
  The latter involves quark-quark-gluon correlations and contains interferences between two light-front Fock sectors, $|qqq\rangle$ and $|qqqg\rangle$, which are usually neglected in the WW approximation. 
  This work provides new nonperturbative input for future experiments at the EicC and EIC on high-twist effects, especially interference effects among multiple partons.
\end{abstract}
\end{frontmatter}

\section{Introduction}
In recent years, the investigation of partonic structures of the nucleon has emerged as a forefront field, attracting significant attention in both experimental and theoretical realms of hadron physics. 
The simplest observables for characterizing these parton structures are the parton distribution functions (PDFs), encoding the nonperturbative structure of a hadron in terms of the distribution of longitudinal momentum fraction $x$ carried by the quarks and gluons~\cite{Jaffe:1983hp,Barone:2001sp,Collins:2011zzd}. 
In contrast to PDFs, transverse-momentum-dependent parton distribution functions (TMDs) are sensitive to the parton's intrinsic transverse motion and spin-dependent correlations, providing a more comprehensive understanding of the parton structures, particularly the transverse structures of hadrons~\cite{Barone:2001sp,Collins:2011zzd,Rogers:2015sqa}. 
These distributions play a crucial role in investigating cross sections of high-energy scattering processes involving hadrons. According to factorization theorems~\cite{Collins:1985ue,Collins:1987pm,Collins:1996fb,Sterman:1995fz,Collins:1998rz,Diehl:2003ny,Collins:2011zzd,Diehl:2011yj,Rogers:2015sqa,Gamberg:2022lju}, PDFs and TMDs, especially for quarks, can be extracted from experiments including deep inelastic scattering (DIS) processes, semi-inclusive deep inelastic scattering (SIDIS) processes and the Drell-Yan experiments, respectively. 

In the theoretical descriptions of high-energy scattering processes, the cross sections are expanded in powers of $1/Q$. 
The contribution of the leading power involves the leading-twist (twist-2) distributions. 
Within the parton model, the twist-2 distributions at leading-order are interpreted as the probability density of finding a parton within a hadron~\cite{Jaffe:1983hp,Barone:2001sp,Collins:2011zzd,Aybat:2011zv}. The subleading power of the $1/Q$ expansion involves the twist-3 distributions. 
Beyond the probabilistic interpretation provided by twist-2 distributions, the twist-3 TMDs and PDFs describe multiparton correlations inside hadrons~\cite{Jaffe:1991kp,Jaffe:1991ra,Efremov:2002qh,AbdulKhalek:2021gbh}.
%
More specifically, the lowest moment of the twist-3 PDF $e(x)$ is related to the mass formation of hadrons~\cite{Hatta:2020iin,Ji:2020baz}. 
The second moment of $e(x)$ is associated with the transverse force and the symmetric and traceless part of the matrix element of the operator $\bar{\psi}D_\mu D_\nu \psi$ between the proton state, where $\psi$ and $D_\mu$ represent the quark field and the covariant derivative, respectively. The matrix element of the operator plays a role in the analysis of DIS and applications of the QCD sum rule method to finite density~\cite{Gubler:2015uza}. 
The twist-3 PDF $g_T(x)$ is related to a transverse color Lorentz force on a quark in a transversely polarized proton~\cite{Burkardt:2008ps,Burger:2021knd}.

While the twist-3 contribution in the cross section is suppressed by $1/Q$, it does not signify that the twist-3 distributions are small. On the contrary, due to the equation of motion (EOM) relations~\cite{Efremov:2002qh,Bacchetta:2006tn,Lorce:2014hxa}, 
the twist-3 distributions are comparable in magnitude to the twist-2 distributions~\cite{Lorce:2014hxa, Pasquini:2018oyz, Bhattacharya:2020cen, Bhattacharya:2021moj, Zhu:2023lst}.
They are measureable in the kinematical region where $Q$ is not large, especially in experiments at the future EicC~\cite{Anderle:2021wcy,Anderle:2021dpv,Zeng:2022lbo,Zeng:2023nnb} and EIC~\cite{AbdulKhalek:2021gbh, Burkert:2022hjz}.
Due to the nonperturbative aspects of QCD and the incomplete understanding of QCD color confinement and chiral symmetry breaking, obtaining these distribution functions directly from first principles poses great challenges.
Twist-3 distributions have been investigated in various phenomenological models and theoretical approaches, for example, the spectator model~\cite{Lu:2014fva,Liu:2021ype}, the diquark spectator approach~\cite{Jakob:1997wg,Mao:2013waa,Mao:2014aoa}, the scalar diquark model~\cite{Lu:2012gu}, the MIT bag model~\cite{Jaffe:1991ra,Signal:1996ct,Avakian:2010br,Lorce:2014hxa}, the constituent parton model~\cite{Lorce:2016ugb,Pasquini:2018oyz}, the perturbative light-front Hamiltonian approaches with a quark target~\cite{Burkardt:2001iy,Kundu:2001pk,Mukherjee:2009uy,Accardi:2009au,Ma:2020kjz}, the covariant model~\cite{Bastami:2020rxn}, the Bethe-Salpeter amplitude approaches~\cite{dePaula:2023ver}, the quark-diquark model~\cite{Sharma:2023wha}, lattice QCD~\cite{Bhattacharya:2020cen, Bhattacharya:2021moj,Bhattacharya:2023nmv}, and the Basis Light-Front Quantization (BLFQ) approach~\cite{Vary:2009gt,Zhu:2023lst,Zhang:2023xfe}.

In this work, we present the twist-3 transverse structures of the proton obtained with the BLFQ framework~\cite{Vary:2009gt}, which represents a nonperturbative approach for solving relativistic many-body bound state problems in light-front quantum field theories.
Previously, this method has demonstrated success in examining the leading-twist TMDs, generalized parton distributions (GPDs) and PDFs of the spin-$0$, spin-$1/2$, and spin-$1$ particles within both QED and QCD.
Specifically, it has been applied to particles such as the electron~\cite{Hu:2020arv}, photon~\cite{Nair:2022evk}, pion~\cite{Lan:2019vui,Lan:2019rba,Zhu:2023lst}, $\rho$-meson~\cite{Kaur:2024iwn}, heavy mesons~\cite{Lan:2019img}, heavy baryons~\cite{Peng:2022lte,Zhu:2023nhl}, and protons~\cite{Mondal:2019jdg,Xu:2021wwj,Hu:2022ctr}. More recently, the BLFQ framework has been extended to successfully investigate subleading twist TMDs of the pion~\cite{Zhu:2023lst} and the TMDs and GPDs of the proton~\cite{Zhang:2023xfe,Liu:2024umn,Yu:2024mxo}.

In the BLFQ framework, we adopt the light-front Hamiltonian and solve for it mass eigenvalues and eigenstates. 
We solve the light-front Hamiltonian in the two Fock sectors $|qqq\rangle$ and $|qqqg\rangle$. 
Our Hamiltonian includes light-front QCD interactions relevant to those Fock components, $3q$ and $3q+g$, and a model confinement in the valence Fock sector $|qqq\rangle$. 
We solve this Hamiltonian and fix the model parameters by fitting the proton mass and electromagnetic properties~\cite{Xu:2023nqv}. 
The resulting eigenstates, \ie, the light-front wavefunctions (LFWFs) with two Fock sectors have been employed to compute various observables of the proton, for example, the twist-2 PDFs~\cite{Xu:2023nqv}, gluon twist-2 GPDs~\cite{Lin:2023ezw}, and T-even gluon leading-twist TMDs~\cite{Yu:2024mxo}. 
In this work, we report a systematic theoretical investigations for twist-3 TMDs of the proton, including the genuine twist-3 TMDs.

\section{Proton LFWFs in the BLFQ framework\label{Sec2}}
In the light-front field theory, the proton's state is obtained as a solution of the light-front stationary Schr\"{o}dinger equation, $P^+P^-|P,\Lambda\rangle=M^2|P,\Lambda\rangle$, where $P^\pm=P^0\pm P^3$~\cite{Kogut:1969xa,Brodsky:1997de}. 
Here\footnote{We adopt a convention distinct from that of Refs.~\cite{Kogut:1969xa,Barone:2001sp}, \ie, the light-front variables are defined by $V^\pm=V^0\pm V^3$, $\vec V_\perp=(V_1,V_2)$. Consequently, the projection operators and phase factors in the definitions of TMD correlators are also different.}, $P^-$ and $P^+$ are the light-front Hamiltonian and the longitudinal momentum of the proton with the mass-squared eigenvalue $M^2$, respectively, and $\Lambda$ represents the proton light-front helicity. 

At the fixed light-front time, $x^+\equiv x^0+x^3$, the nucleon state can be schematically expressed in terms of various quark, antiquark, and gluon Fock components,
\begin{equation}
|P,\Lambda\rangle=\psi_{(qqq)}|qqq\rangle+\psi_{(qqqg)}|qqqg\rangle+\psi_{(qqqq\bar{q})}|qqqq\bar{q}\rangle+\cdots,\label{psi0}
\end{equation}
where $\psi_{(\cdots)}$ is the LFWF encoding the structural information of the system in the different Fock sectors $|\cdots\rangle$. 
For numerical calculations, one needs to truncate the infinite Fock sector expansion in Eq.~(\ref{psi0}) to a finite Fock space. 
At the model scale, we consider only the valence-quark and valence-quark-gluon Fock components, $3q$ and $3q+g$, respectively. 
Currently, we adopt an effective light-front Hamiltonian, $P^-=P^-_{\mathrm{QCD}}+P_{\mathrm{C}}^-$, instead of the complete light-front QCD Hamiltonian $P_{\mathrm{QCD}}^-$~\cite{Xu:2023nqv}. 

In the light-front gauge $A^+=0$, the light-front QCD Hamiltonian with one dynamical gluon~\cite{Lan:2021wok,Xu:2023nqv} is
\begin{align}
    P_{\rm QCD}^-= &\int \mathrm{d}x^- \mathrm{d}^2 x^{\perp} \Big\{\frac{1}{2}\bar{\psi}\gamma^+\frac{m_{0}^2+(i\partial^\perp)^2}{i\partial^+}\psi\nonumber\\
    &-\frac{1}{2}A_a^i\left[m_g^2+(i\partial^\perp)^2\right] A^i_a +g_s\bar{\psi}\gamma_{\mu}T^aA_a^{\mu}\psi \nonumber\\
    &+ \frac{1}{2}g_s^2\bar{\psi}\gamma^+T^a\psi\frac{1}{(i\partial^+)^2}\bar{\psi}\gamma^+T^a\psi \Big\},
\end{align}
where $x^-$ and $x^\perp$ are the longitudinal and transverse coordinates, respectively.
$\psi$ and $A^\mu$ represent the quark and gluon fields, respectively. 
$\gamma^\mu$ is the Dirac matrix.
$T^a$ is the half Gell-Mann matrix, $T^a=\lambda^a/2$. 
$g_s$ is the coupling constant. 
$m_0$ and $m_g$ denote the bare quark mass and the model gluon mass, respectively. 
Although the gluon mass is zero in QCD, we allow for a phenomenological gluon mass to fit the nucleon form factors. 
There are also effective gluon masses appearing in the Dyson-Schwinger equations (DSEs) approach~\cite{Cornwall:1981zr,Alkofer:2000wg} and other theoretical approaches~\cite{Deur:2016tte}.
To account for the quark mass correction from a higher Fock sector, we introduce a mass counterterm, $\delta m_q=m_0-m_q$, for the quark in the leading Fock sector, where $m_q$ represents the renormalized quark mass. 
Following Ref.~\cite{Glazek:1992aq}, we introduce an independent quark mass $m_f$ in the vertex interaction. 

The additional term, $P^-_\mathrm{C}$, is a confining potential exclusively in the valence Fock sector. It is described by the following form~\cite{Lan:2021wok,Li:2015zda},
\begin{align}
    P^+P^-_{\mathrm{C}}=\frac{\kappa^4}{2}\sum_{i\neq j}\Big\{\vec{r}_{ij\perp}^2-\frac{\partial_{x_i}(x_ix_j\partial_{x_j})}{(m_i+m_j)^2}\Big\},
\end{align}
where the relative coordinate is $\vec{r}_{ij\perp}=\sqrt{x_ix_j}(\vec{r}_{i\perp}-\vec{r}_{j\perp})$ related to the holographic variable~\cite{Brodsky:2014yha}. The parameter $\kappa$ is the strength of the confinement, and $\partial_{x}\equiv({\partial}/\partial)_{r_{ij\perp}}$.
The explicit confinement is omitted in the $|qqqg\rangle$ sector with the expectation that the limited basis in the transverse direction (see below) and the presence of the massive gluon model the effects of confinement.

Under the BLFQ framework~\cite{Vary:2009gt,Xu:2023nqv}, the Fock-sector basis states are direct products of the Fock particle states $|\alpha\rangle=\otimes_i|\alpha_i\rangle$. 
In the longitudinal direction, the $i^{\mathrm{th}}$ Fock particle is confined to a one-dimensional box of length $2L$, with antiperiodic (periodic) boundary condition for fermions (bosons). 
Specifically, the longitudinal momentum is discretized as $p_i^+=\frac{2\pi}{L}k_i$, where $k_i$ is a half-integer (integer) for fermions (bosons). 
Here, we omit the zero mode of bosons. In the transverse plane, we consider the two-dimensional harmonic oscillator (2-D HO) basis function, $\Phi_{n_im_i}(\vec{p}_{\perp i};b)$, with a scale parameter $b$, where $n_i$ and $m_i$ are the radial and angular quantum numbers, respectively. $\vec{p}_{\perp i}$ is the transverse momentum.
Each Fock-particle basis state $|\alpha_i\rangle$ is characterized by the three quantum numbers mentioned above and the light-front helicity $\lambda_i$, \ie, $|\alpha_i\rangle=|k_i,n_i,m_i,\lambda_i\rangle$. 
All the Fock-sector basis states have the same total angular momentum projection $\Lambda$ such that $\sum_i(\lambda_i+m_i)=\Lambda$.

We introduce two truncation parameters to limit the infinite bases in the longitudinal and transverse directions. 
The longitudinal truncation is represented by $K=\sum_i k_i$, parameterizing the total longitudinal momentum $P^+$. 
Thus, for the $i^{\mathrm{th}}$ Fock particle, the longitudinal momentum fraction is defined as $x_i=p^+/P^+=k_i/K$. The transverse truncation $N_{\rm max}$ restricts the 2-D HO basis states of many Fock particles in the Fock sectors, $\sum_i(2n_i+|m_i|+1)\le N_{\rm max}$. 
The transverse truncation implicitly serves as the infrared (IR) and ultraviolet (UV) cutoffs in momentum space, with $\lambda_{\rm IR} \simeq b/\sqrt{N_{\rm max}}$ and $\Lambda_{\rm UV} \simeq b\sqrt{N_{\rm max}}$~\cite{Wiecki:2014ola,Zhao:2014xaa}.

Following Ref.~\cite{Xu:2023nqv}, we restrict our calculation within the truncation $\{N_{\mathrm{max}}$, $K\}$ = $\{9$, $16.5\}$.
We select the HO scale parameter $b=0.7\;\mathrm{GeV}$, the UV cutoff for the instantaneous interaction $b_{\mathrm{inst}}=3\;\mathrm{GeV}$, and adopt the model parameters $\{m_u$, $m_d$, $m_g$, $\kappa$, $m_f$, $g_s\}$ = $\{0.31$, $0.25$, $0.50$, $0.54$, $1.80$, $2.40\}$ with units of $\mathrm{GeV}$ except for $g_s$, which are fitted to the proton mass and electromagnetic properties. 
After solving the light-front Schr\"odinger equation, we obtain the LFWFs, 
\begin{align}
    &\Psi_{\mathcal{N},\{\lambda_{i}\}}^{\Lambda}({\{x_{i}, \vec{p}_{i \perp}\}})\nonumber\\
    &=\sum_{\{n_i,m_i\}}\psi^{\Lambda}_{\mathcal{N}}(\{\alpha_i\}) \prod_{i}^{\mathcal{N}} \phi_{n_{i},m_i}\left(\vec{p}_{i \perp} ; b\right),\label{eq:LFWF}
\end{align}
where $\psi^{\Lambda}_{\mathcal{N}=3}(\{\alpha_i\})$ and $\psi^{\Lambda}_{\mathcal{N}=4}(\{\alpha_i\})$ are the components of the eigenvectors associated with the Fock sectors $|qqq\rangle$ and $|qqqg\rangle$, respectively.
Note that $|qqqg\rangle$ has two color-singlet states, necessitating one additional label for specification.
At the model scale, the probabilities of the proton being in the Fock states $|qqq\rangle$ and $|qqqg\rangle$ are $44\%$ and $56\%$ respectively.

\section{Transverse-momentum-dependent parton distributions}
For spin-$1/2$ baryons, the quark TMDs are parameterized through the following TMD correlation function~\cite{Goeke:2005hb,Bacchetta:2006tn,Meissner:2009ww},
\begin{align}
    &\Phi_{ij}\left(x,  k_{\perp};S\right)=\int \frac{\mathrm{d} z^{-} \mathrm{d}^2 z_{\perp}}{2(2 \pi)^{3}} e^{i k \cdot z}\nonumber\\
  &\times\left.\left\langle P, S\right|{\bar{\psi}}_j(0) \mathcal{U}^{n_-}_{(0,+\infty)}\mathcal{U}^{n_-}_{(+\infty,z)} {\psi}_i(z)\left| P, S\right\rangle\right|_{z^{+}=0},\label{eq:TMD_correlation_function}
\end{align}
with $x=k^+/P^+$. $\psi$ is the quark field. $\mathcal{U}^{n_-}_{(0,+\infty)}$ and $\mathcal{U}^{n_-}_{(+\infty,z)}$ are the gauge links ensuring the gauge invariance of the bilocal quark field operators~\cite{Buffing:2011mj, Buffing:2013dxa, Buffing:2013kca, Bacchetta:2020vty, Bacchetta:2024fci}. $|P,S\rangle$ denotes the bound state of the baryon with mass $M$, spin $S$, and four-momentum $P$ where the transverse momentum is zero $\vec P_\perp=\vec{0}$~\cite{Collins:1992kk}. 

By analyzing parity, charge conjugation, hermiticity invariance, and twist expansion, one can parameterize the TMD correlation function into the twist-2, twist-3, and twist-4 TMDs. 
In this work, we specifically focus on the twist-3 level. 
The Dirac matrices, $\gamma^+$, $\gamma^+\gamma_5$, and $i\sigma_{j+}\gamma^5$, project the TMD correlation function into eight twist-2 TMDs~\cite{Goeke:2005hb,Bacchetta:2006tn,Meissner:2009ww},
\begin{align}
  \Phi^{\left[\gamma^{+}\right]}&\left(x,  k_{\perp} ; S\right)=f_{1}-\frac{\epsilon_{\perp}^{i j} k_\perp^{i} S_\perp^{j}}{M} f_{1 T}^{\perp}, \label{eq:TMDs1}\\
  \Phi^{\left[\gamma^{+} \gamma^{5}\right]}&\left(x,  k_{\perp} ; S\right)=S^{3} g_{1 L}+\frac{ \vec k_{\perp} \cdot \vec S_\perp}{M} g_{1 T}, \label{eq:TMDs2}\\
  \Phi^{\left[i \sigma^{j+} \gamma^{5}\right]}&\left(x,  k_{\perp} ; S\right)=S_\perp^{j} h_{1}+S^{3} \frac{k_\perp^{j}}{M} h_{1 L}^{\perp}\nonumber\\
  &+S_\perp^{i} \frac{2 k_\perp^{i} k_\perp^{j}-(\vec k_{\perp})^{2} \delta^{i j}}{2 M^{2}} h_{1 T}^{\perp}+\frac{\epsilon_{\perp}^{j i} k_\perp^{i}}{M} h_{1}^{\perp}.\label{eq:TMDs3}
\end{align}
The Dirac matrices, $\mathbbm{1}$, $i\gamma_5$, $\gamma^i$, $\gamma^i\gamma_5$, $i\sigma^{ij}\gamma_5$ and $i\sigma^{+-}\gamma_5$, project the TMD correlation function into sixteen twist-3 TMDs~\cite{Goeke:2005hb,Bacchetta:2006tn,Meissner:2009ww},
\begin{align}
  \Phi^{[\mathbbm{1}]} &\left(x,  k_{\perp} ; S\right)=\frac{M}{P^{+}}\left[e-\frac{\epsilon_{\perp}^{ij} k_{\perp }^i S_{\perp}^j}{M} \textcolor{black}{e_{T}^{\perp}}\right],\label{eq:e} \\
  \Phi^{\left[i \gamma_{5}\right]} &\left(x,  k_{\perp} ; S\right)=\frac{M}{P^{+}}\left[S^3_{} \textcolor{black}{e_{L}}+\frac{\vec k_{\perp} \cdot\vec S_{\perp}}{M} \textcolor{black}{e_{T}}\right], \\
  \Phi^{\left[\gamma^{i}\right]} &\left(x,  k_{\perp} ; S\right)=\frac{M}{P^{+}}\Bigg[\epsilon_{\perp}^{ij} S_{\perp}^j \textcolor{black}{f_{T}}+S_{}^3 \frac{\epsilon_{\perp}^{ij} k_{\perp}^j}{M} \textcolor{black}{f_{L}^{\perp}}\nonumber\\
  &-\frac{2 k_{\perp}^{i} k_{\perp}^{j}- (\vec k_{\perp})^{2} \delta^{ij}}{2M^{2}} \epsilon_{\perp}^{jk} S_{\perp}^{k} \textcolor{black}{f_{T}^{\perp}}+\frac{k_{\perp}^{i}}{M} f^{\perp}\Bigg], \\
  \Phi^{\left[\gamma^{i} \gamma_{5}\right]} &\left(x,  k_{\perp} ; S\right)=\frac{M}{P^{+}}\Bigg[S_{\perp}^{i} g_{T}+S^3 \frac{k_{\perp}^{i}}{M} g_{L}^{\perp}\nonumber\\
  &+S_{\perp}^j\frac{2k_{\perp}^{i} k_{\perp}^{j}-(\vec k_{\perp})^{2} \delta^{ij}}{2M^{2}}  g_{T}^{\perp}+\frac{\epsilon_{\perp}^{ij} k_{\perp }^j}{M} \textcolor{black}{g^{\perp}}\Bigg], \\
  \Phi^{\left[i \sigma^{ij } \gamma_{5}\right]} &\left(x,  k_{\perp} ; S\right)=\frac{M}{P^{+}}\left[\frac{S_{\perp}^{i} k_{\perp}^{j}-k_{\perp}^{i} S_{\perp}^{j}}{M} h_{T}^{\perp}-\epsilon_{\perp}^{ij} \textcolor{black}{h}\right], \\
  \Phi^{\left[i \sigma^{+-} \gamma_{5}\right]} &\left(x,  k_{\perp} ; S\right)=\frac{M}{P^{+}}\left[S_{}^3 h_{L}+\frac{\vec k_{\perp} \cdot \vec S_{\perp}}{M} h_{T}\right]\label{eq:h_L},
\end{align}
where $\Phi^{[\Gamma]}=\frac{1}{2}\mathrm{Tr}[\Phi_{}\Gamma_{}]$.
$i$ and $j$ are the transverse indices. 
$\epsilon_T^{12}=-\epsilon_T^{21}=1$ and $\epsilon_T^{11}=\epsilon_T^{22}=0$. $S^3$ represents the light-front helicity of the proton, while $S_\perp$ denotes the transverse component of the polarization vector of the proton. Among them, the twist-2 T-odd TMDs include $f_{1T}^\perp$ and $h_1^\perp$, and the twist-3 T-odd TMDs include $e_T^\perp$, $e_L$, $e_T$, $f_T$, $f_L^\perp$, $f_T^\perp$, $g^\perp$, and $h$. The others are T-even TMDs~\cite{Bacchetta:2006tn,Meissner:2009ww}. These T-odd TMDs are non-zero when considering the nontrivial contributions from the gauge links~\cite{Lorce:2016ugb}. Note that there is a suppression with the factor $M/P^+$ before the square bracket in Eqs.~(\ref{eq:e}-\ref{eq:h_L}). But it does not mean that the value of twist-3 TMDs is smaller than twist-2 TMDs. On the contrary, the twist-3 TMDs in square brackets are often large enough so that the magnitude of the l.h.s. twist-3 TMD correlation functions is comparable to the magnitude of the twist-2 TMD correlation functions.

While there exist sixteen twist-3 TMDs for spin-1/2 baryons, they are not independent~\cite{Efremov:2002qh,Bacchetta:2006tn,Lorce:2014hxa}. 
The twist-3 TMDs can be decomposed into contributions from the twist-2 TMDs, the quark-quark-gluon correlation functions and the singular terms, $\delta(x)$, as dictated by the equation of motion (EOM) relation~\cite{Efremov:2002qh}. 
In the light-front field theory, the quark field is decomposed into a ``good'' $\psi_+$, and a ``bad'' $\psi_-$ component ($\psi_\pm=\frac{1}{4}\gamma^\mp\gamma^\pm\psi$). 
The bad component is expressed in terms of the good component and the transverse gauge field, $A_{\perp j}$, under the light-cone gauge $A^+=0$~\cite{Brodsky:1997de},
\begin{align}
    \psi_-(z)=\frac{\gamma^+}{2i\partial^+}[i(\partial_j-ig_sA_{\perp j}(z))\gamma_j+m]\psi_+(z),
\end{align}
where $m$ represents the quark mass, and $j=1,2$.

For the leading twist TMDs, the Dirac matrices project the fermion bilocal operators into $\bar{\psi}_+\psi_+$ configuration. 
One finds that the operator combinations in the leading twist TMD correlation functions are just densities or differences of densities~\cite{Barone:2001sp,Collins:2011zzd,Zhu:2023nhl}.  
However, in the case of the subleading twist TMDs, the Dirac matrices project the fermion bilocal operators into $\bar{\psi}_-\psi_++\bar{\psi}_+\psi_-$ configurations, \ie, the twist-3 distributions do not have probabilistic interpretations. 
They describe multi-parton distributions corresponding to the interference between LFWFs in different Fock sectors and reflect the physics of quark-quark-gluon correlations, specifically between scattering from a coherent quark-gluon pair and from a single quark~\cite{Jaffe:1991kp,Mulders:1995dh,Jaffe:1991ra,AbdulKhalek:2021gbh}.

Through the EOM relations (see below), the twist-3 TMDs can be expressed in terms of the twist-2 TMDs and genuine twist-3 TMDs corresponding to quark-quark-gluon interaction terms. It is convenient to introduce the quark-quark-gluon correlation function~\cite{Mulders:1995dh,Bacchetta:2006tn},
\begin{align}
    \tilde{\Phi}^{i}_A(x, k_\perp;S)&=\Phi^{i}_{D}(x,k_\perp;S)-k_\perp^i\Phi(x,k_\perp;S),
    \label{eq:definition_Phi_A}
\end{align}
where the correlation function $\Phi$ is defined in Eq.~(\ref{eq:TMD_correlation_function}). The correlation function related to the covariant derivative is defined as~\cite{Mulders:1995dh,Bacchetta:2006tn},
\begin{align}
   (\Phi^\mu_D)_{ij}&(x,k_\perp;S) =\int\frac{\mathrm{d}z^-\mathrm{d}^2z_\perp}{2(2\pi)^3}e^{ik\cdot z}\langle P, S|{\bar{\psi}}_j(0)\mathcal{U}^{n_-}_{(0,+\infty)}\nonumber\\
    &\times\mathcal{U}^{n_-}_{(+\infty,z)} iD^\mu(z) {\psi}_i(z)| P, S\rangle|_{z^{+}=0},\label{TMD}
\end{align}
with $iD^\mu=i\partial ^\mu +g_s A^\mu$. Note that obtaining nonvanishing quark-quark-gluon correlation functions defined in Eq.~(\ref{eq:definition_Phi_A}) requires considering a higher Fock component with a dynamical gluon.

Like the twist-2 and twist-3 TMDs, by analyzing symmetries, one can also parameterize the quark-quark-gluon correlator into sixteen genuine twist-3 TMDs~\cite{Bacchetta:2006tn},
\begin{align}
    \frac{1}{ M x} &\tilde{\Phi}_{A i}^{[\sigma^{i+}]}(x,k_\perp;S)=\textcolor{black}{\tilde{h}}+i \tilde{e}+\frac{\epsilon_\perp^{jk} k_{\perp }^j S_{\perp }^k}{M}(\tilde{h}_T^{\perp}-i \textcolor{black}{\tilde{e}_T^{\perp}}),\label{eq:gTMD1} \\
    \frac{1}{ M x} &\tilde{\Phi}_{A i}^{[i \sigma^{i+} \gamma_5]}(x,k_\perp;S)=S^3(\tilde{h}_L+i \textcolor{black}{\tilde{e}_L})+\frac{\vec k_{\perp} \cdot \vec S_\perp}{M}(\tilde{h}_T+i \textcolor{black}{\tilde{e}_T}), \\
    \frac{1}{ M x} &\tilde{\Phi}_{A j}^{[(g_\perp^{ij }+i \epsilon_\perp^{ij } \gamma_5) \gamma^{+}]}(x,k_\perp;S)=\frac{k_{\perp}^i}{M}(\tilde{f}^{\perp}-i \textcolor{black}{\tilde{g}^{\perp}})\nonumber\\
    &+\epsilon_\perp^{ij} S_{\perp}^j(\textcolor{black}{\tilde{f}_T}+i \tilde{g}_T)+S^3 \frac{\epsilon_\perp^{ij} k_{\perp}^j}{M}(\textcolor{black}{\tilde{f}_L^{\perp}}+i \tilde{g}_L^{\perp})\nonumber\\
    &-\frac{2k_{\perp}^i k_{\perp}^j- (\vec{k}_{\perp})^2 \delta^{ij}}{2M^2} \epsilon_{\perp}^{jk} S_\perp^k(\textcolor{black}{\tilde{f}_T^{\perp}}+i \tilde{g}_T^{\perp}),\label{eq:gTMD3} 
  \end{align}
where $\tilde{\Phi}_{Aj}^{[\Gamma]}=\frac{1}{2}\mathrm{Tr}[\tilde{\Phi}_{Aj}\Gamma_{}]$ and $\tilde{\Phi}_{Aj}=\tilde{\Phi}_A^i g_{\perp ij}$. There are eight T-odd genuine twist-3 TMDs, $\tilde{h}$, $\tilde{e}_T^\perp$, $\tilde{e}_L$, $\tilde{e}_T$, $\tilde{g}^\perp$, $\tilde{f}_T$, $\tilde{f}^\perp_L$, and $\tilde{f}_T^\perp$. The rest are T-even~\cite{Bacchetta:2006tn}.

Here, we neglect the nontrivial impact of gauge links and employ unit matrix approximations for the gauge links, $\mathcal{U}^{n_-}_{(0,+\infty)}\approx \mathbbm{1}$ and $\mathcal{U}^{n_-}_{(+\infty,z)}\approx \mathbbm{1}$. Consequently, only the T-even TMDs survive~\cite{Lorce:2016ugb}. 
Furthermore, we neglect the zero modes, leading to the disappearance of singular terms, $\delta(x)$, in the EOM relation~\cite{Efremov:2002qh,Bacchetta:2006tn,Lorce:2014hxa}.
With some algebra, the EOM relations relating T-even TMDs from Eqs.~(\ref{eq:TMDs1}-\ref{eq:TMDs3}), Eqs.~(\ref{eq:e}-\ref{eq:h_L}) and Eqs.~(\ref{eq:gTMD1}-\ref{eq:gTMD3}) can be derived as~\cite{Bacchetta:2006tn,Lorce:2014hxa,Pasquini:2018oyz},
\begin{align}
    e &=\tilde{e}+\frac{m}{M} \frac{f_{1}}{x},\label{eq:Eom1} \\
    f^{\perp} &=\tilde{f}^{\perp}+\frac{f_{1}}{x}, \\
    g_{T}^{\perp} &=\tilde{g}_{T}^{\perp}+\frac{g_{1 T}}{x}+\frac{m}{M} \frac{h_{1 T}^{\perp}}{x}, \\
    g_{T}^{} &=\tilde{g}_{T}^{}+\frac{(\vec{k}_\perp)^2}{2M^2}\frac{g_{1T}}{x}+\frac{m}{M} \frac{h_{1}}{x} ,\\
    g_{L}^{\perp}&=\tilde{g}_{L}^{\perp}+\frac{g_{1 L}}{x}+\frac{m}{M} \frac{h_{1 L}^{\perp}}{x}, \\
    h_{L} &=\tilde{h}_{L}+\frac{m}{M} \frac{g_{1 L}}{x}-\frac{(\vec{k}_\perp)^2}{M^{2}} \frac{h_{1 L}^{\perp}}{x}, \\
    h_{T}&=\tilde{h}_{T}-\frac{h_1}{x}-\frac{(\vec k_\perp)^2}{2M^2}\frac{h_{1T}^\perp}{x}+\frac{m}{M}\frac{g_{1T}}{x},\\
    h_{T}^{\perp}&=\tilde{h}_{T}^{\perp}+\frac{h_1}{x}-\frac{(\vec k_\perp)^2}{2M^2}\frac{h_{1 T}^\perp}{x},\label{eq:Eom8} 
\end{align}
where all the terms on the l.h.s represent the twist-3 T-even TMDs, the tilde terms denote the genuine twist-3 T-even TMDs, and the remaining terms correspond to the twist-2 T-even TMDs. In the Wandzura-Wilczek (WW) approximation~\cite{Bastami:2018xqd}, all the tilde terms and the terms involving the quark mass $m$ are ignored.

In this study, we compute the T-even TMDs of the proton from LFWFs in different Fock sectors, $|uud\rangle$ and $|uudg\rangle$. The LFWFs from two different Fock sectors are simultaneously obtained by diagonalizing the light-front QCD Hamiltonian within the BLFQ framework. The detailed overlap expressions of all T-even TMDs are presented in the appendix.

\section{Numerical results}
With the model parameters for the basis truncation $\{N_{\rm{max}} ,K \} = \{9, 16.5\}$ mentioned at the end of Sec.~\ref{Sec2}, we solve the light-front stationary Schr\"odinger equation, yielding the two Fock-state LFWFs of the proton with the mass $M = 0.956\; \mathrm{GeV}$. 
Subsequently, we utilize the resulting LFWFs along with the expressions in Eqs. (\ref{eq:TMDoverlap}) and (\ref{eq:gTMDoverlap}) to compute the light-front helicity amplitudes. Employing the amplitudes and Eqs.~(\ref{eq:f_1}-\ref{eq:gtildeTperp}), we obtain the twist-2 TMDs and the genuine twist-3 TMDs without evolution effects or gauge links for the valence quarks in the proton. 
We then employ the EOM relations, as given in Eqs.~(\ref{eq:Eom1}-\ref{eq:Eom8}), to determine the total twist-3 TMDs. 
In this work, we mainly discuss the twist-3 structures of the proton.

\subsection{Genuine twist-3 TMDs and twist-3 TMDs}
\begin{figure*}[h]
  \centering
  \includegraphics[width=0.6\textwidth]{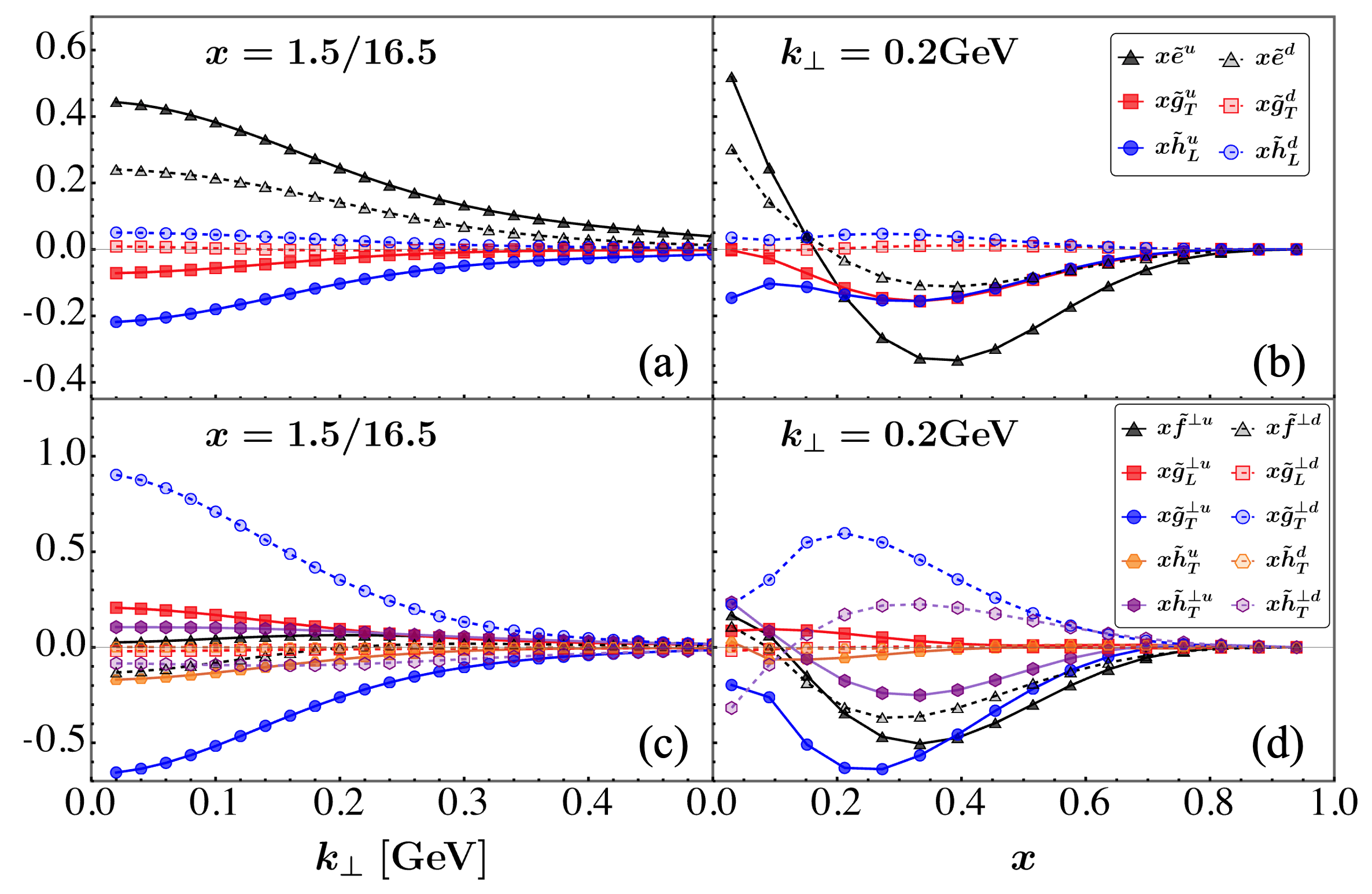}
  \caption{Results for the genuine twist-3 $x$TMDs obtained within the BLFQ framework. The solid (dashed) lines correspond to the distributions of the $u$ ($d$) quark. Panels (a) and (c) present the TMDs as functions of $k_\perp$, while panels (b) and (d) present the TMDs as functions of $x$. 
  Specifically, the TMDs in (a) and (b) provide the genuine twist-3 PDFs upon integration over $k_\perp$.}
  \label{fig:g_tw-3TMD}
\end{figure*}

Figure~\ref{fig:g_tw-3TMD} presents the genuine twist-3 TMDs.
The solid (dashed) lines correspond to the distributions of the $u$ ($d$) quark. 
Panels (a) and (c) in Fig.~\ref{fig:g_tw-3TMD} present the TMDs as functions of $k_\perp$ at fixed longitudinal momentum fraction $x=1.5/16.5$. 
Panels (b) and (d) in Fig.~\ref{fig:g_tw-3TMD} present the TMDs as functions of $x$ at fixed transverse momentum $k_\perp=0.2$ GeV.  
The TMDs in panels (a) and (b) provide the genuine twist-3 PDFs upon integration over $k_\perp$.
%
%

Analyzing the structures of operators in the twist-2 correlators reveals that twist-2 TMDs have a probabilistic interpretation~\cite{Jaffe:1983hp,Barone:2001sp,Collins:2011zzd,Tangerman:1994eh,Zhu:2023nhl}. In contrast, there is no such probabilistic interpretation for genuine twist-3 TMDs. 
%
They correspond to polarized or unpolarized interference distributions~\cite{Mulders:1995dh}.
For genuine T-even twist-3 TMDs, $\tilde{e}$ and $\tilde{f}^\perp$ correspond to unpolarized interference distributions, while others correspond to polarized interference distributions. Specifically, based on Jaffe–Ji classification~\cite{Barone:2001sp}, the notation $g$ ($h$) refers to the longitudinally (transversely) polarized struck quark, and the subscript L (T) refers to the longitudinal (transverse) polarization of the proton. 
In Fig.~\ref{fig:g_tw-3TMD}, our results show that the polarized distributions of $u$ and $d$ quarks of the same type exhibit opposite signs. 
By analyzing the spin structures of the proton in the quark model, one can find that the polarization directions of the $u$ and $d$ quarks are different~\cite{Gell-Mann:1964ewy,Zhu:2023nhl}. Specifically, the $u$ quark tends to be polarized in the same direction as the proton, while the $d$ quark tends to be polarized in the opposite direction.
As a result, their polarized distributions are expected to have opposite signs.
%
%
In addition, as shown in panels (a) and (c), our results show that the magnitude of the genuine twist-3 TMDs decreases as $k_\perp$ increases. 
%

\begin{figure*}[h]
  \centering
  \includegraphics[width=0.6\textwidth]{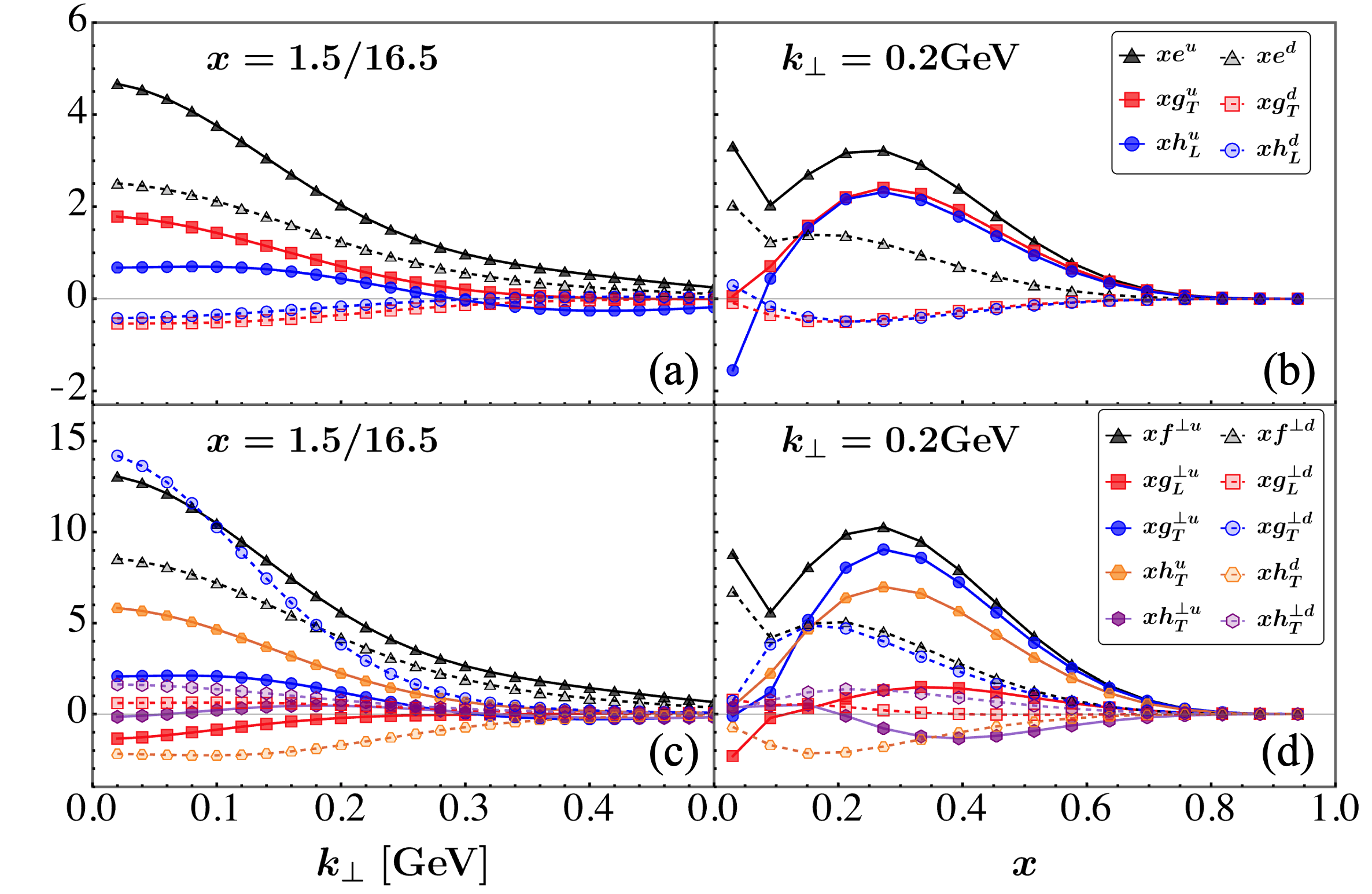}  
  \caption{Results for the twist-3 $x$TMDs obtained within the BLFQ framework. The solid (dashed) lines correspond to the distributions of the $u$ ($d$) quark. Panels (a) and (c) present the $x$TMDs as functions of $k_\perp$, while panels (b) and (d) present the $x$TMDs as functions of $x$. Specifically, the TMDs in (a) and (b) provide the twist-3 PDFs upon integration over $k_\perp$. }
  \label{fig:tw-3TMD}
\end{figure*}
Figure~\ref{fig:tw-3TMD} shows the total twist-3 TMDs. 
The solid (dashed) lines represent the distributions of the $u$ ($d$) quark. 
Panels (a) and (c) in Fig.~\ref{fig:tw-3TMD} display the TMDs as functions of $k_\perp$ at fixed longitudinal momentum fraction $x=1.5/16.5$. 
Panels (b) and (d) in Fig.~\ref{fig:tw-3TMD} depict the TMDs as functions of $x$ at fixed transverse momentum $k_\perp=0.2$ GeV.  
The TMDs in panels (a) and (b) provide the twist-3 PDFs upon integration over $k_\perp$.

Like the genuine twist-3 distribution functions, the twist-3 TMDs have no probability interpretation as mentioned before. The twist-3 distributions are multi-parton distributions involving the interference between different Fock sectors, reflecting the physics of quark-quark-gluon correlations.
By analyzing the parameterized structures of twist-3 TMDs, given in Eqs.~(\ref{eq:e}-\ref{eq:h_L}), the twist-3 distributions $e$ and $f^\perp$ describe unpolarized distributions, \ie, the unpolarized-parton interferences in an unpolarized hadron.
On the other hand, other twist-3 distributions describe polarized distributions, including either unpolarized-parton interferences within a polarized hadron, polarized-parton interferences within an unpolarized hadron, or polarized-parton interferences within a polarized hadron.
Our results show that the amplitude of the unpolarized twist-3 TMDs $e$ and $f^{\perp }$ of the $u$ quark is twice as large as that of the $d$ quark. This arises from the presence of twice as many valence $u$ quarks in a proton compared to the $d$ quark.
The other polarized twist-3 TMDs of $u$ and $d$ quarks of the same type exhibit similar trends but with opposite signs. 
%
However, $g_T^{\perp u}$ and $g_T^{\perp d}$ are exceptions and exhibit positive distributions, showing no sign difference between them.

The positivity bounds~\cite{Soffer:1994ww,Bacchetta:1999kz}, commonly known as Soffer-type bounds, play an important role in validating model calculations for twist-2 TMDs that constitute a significant portion of the twist-3 TMDs.
These bounds, which are universal and model-independent, are established through the examination of the positivity of the spin-density matrix of quarks. Our results satisfy these bounds, implying consistency for the BLFQ calculations. 
In the previous investigation focused on the proton's valence components $3q$, we found that all the twist-2 TMDs in the BLFQ calculations are independent of each other~\cite{Hu:2022ctr}.
This conclusion, not universally present in some model calculations, signifies an essential characteristic of our approach~\cite{Avakian:2010br,Lorce:2011zta}.
In this work, we increase the number of Fock sectors, yielding more independent helicity amplitudes. Therefore, the independence property applies to all the twist-2 and twist-3 distributions.

\subsection{PDFs and sum rules}

\begin{figure}[h]
    \centering
    \includegraphics[width=0.4\textwidth]{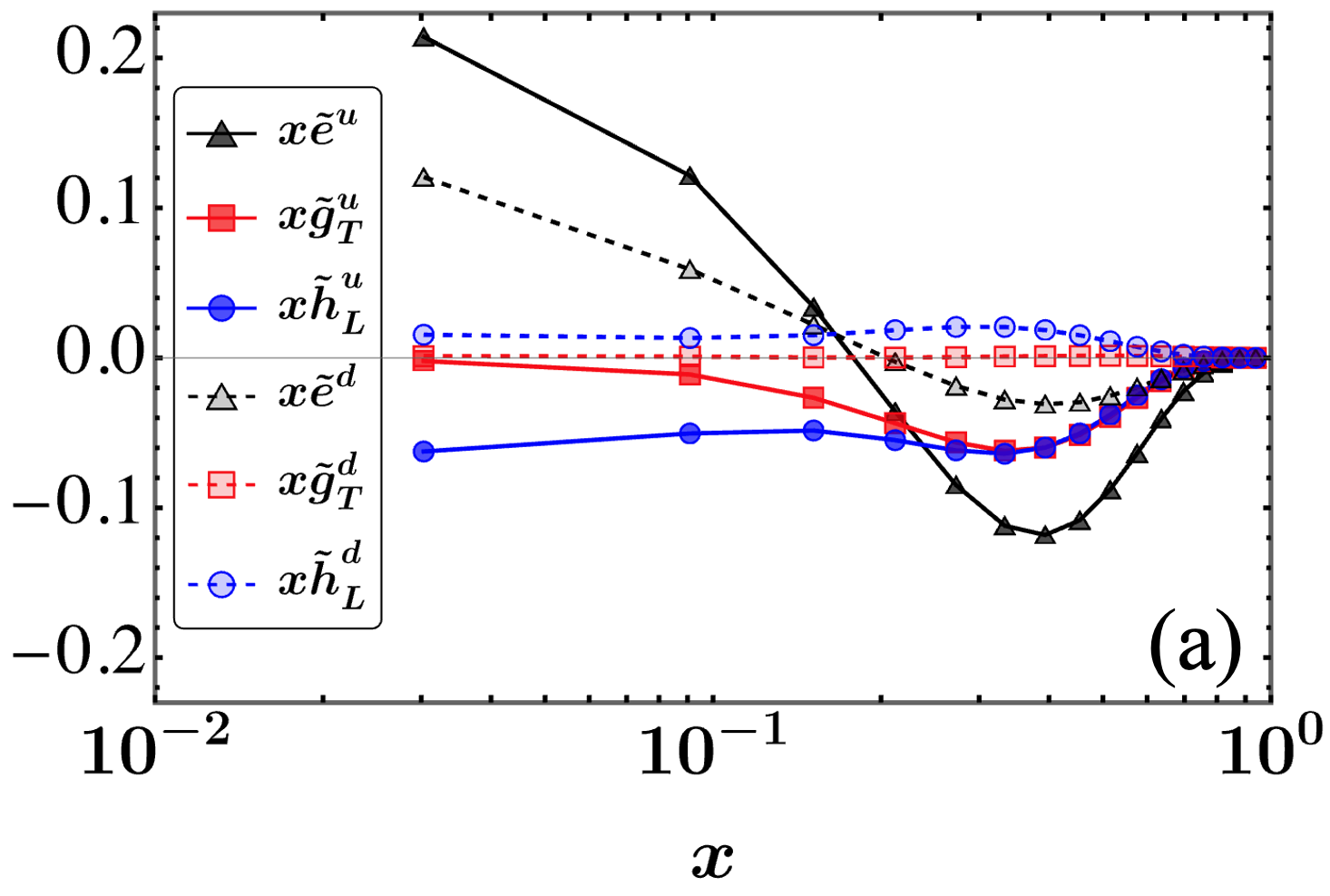} 
    \includegraphics[width=0.4\textwidth]{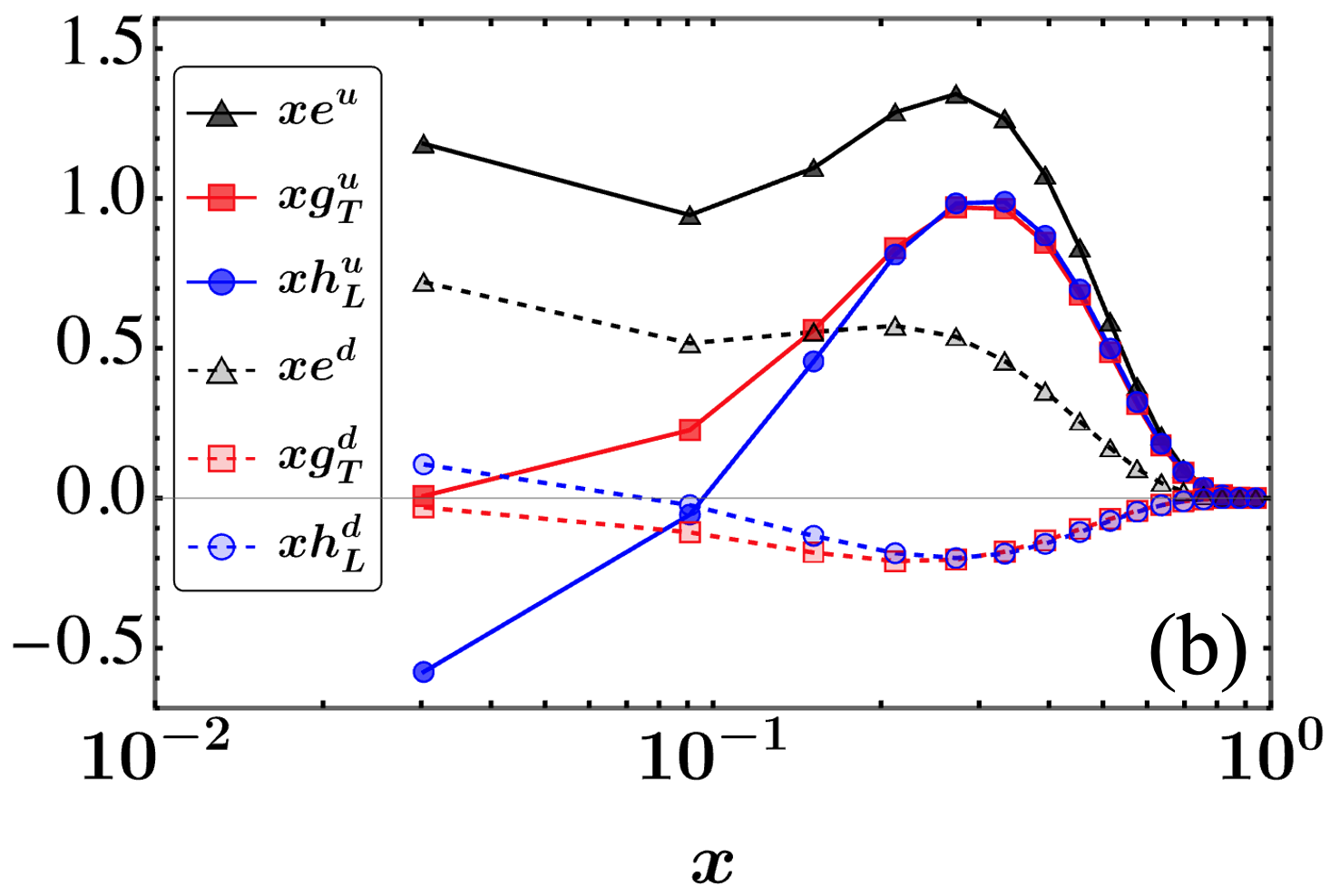} 
    \caption{Results for (a) the genuine twist-3 $x$PDFs and (b) the twist-3 $x$PDFs obtained under the BLFQ framework. The solid (dashed) lines in panels (a) and (b) correspond to the distributions of the $u$ ($d$) quark.}
    \label{fig:xpdf}
\end{figure}

Following Ref.~\cite{Xu:2023nqv}, we determine the initial scale $\mu_0$ by requiring the initial unpolarized PDF $f_1(x)$ after evolution to generate the total first moments of the valence quarks PDFs from the global QCD analysis, $\langle x\rangle_u+\langle x\rangle_d = 0.374$ at $10\; \rm{GeV}^2$~\cite{deTeramond:2018ecg}. This yields $\mu_0^2=0.24\pm0.01\;\mathrm{GeV}^2$.
At the model scale $\mu_0$, integrating TMDs over the transverse momentum $k_\perp$, we obtain the corresponding PDFs. 
We emphasize that the relationship between the integrated TMDs and collinear PDFs only holds under the standard collinear factorization. 
The relationship between the full pQCD treatment of factorization and parton-model intuition has remained much less clear in TMD factorization than in the collinear factorization, especially after accounting for evolution effects, as the two kinds of distributions decouple at high energy scales due to different evolution equations~\cite{Collins:2011zzd,Aybat:2011zv, Echevarria:2011epo}.
Figure~\ref{fig:xpdf} shows the genuine twist-3 $x$PDFs and the twist-3 $x$PDFs at the initial scale. 
Panel (a) in Fig.~\ref{fig:xpdf} shows that the genuine twist-3 $x$PDFs $x\tilde{e}^u(x)$ and $x\tilde{e}^d(x)$ exhibit an obvious sign-changing pattern, being positive for $x\textless 0.2$ and negative for $x\textgreater 0.2$. As $x$ decreases, the values of $x\tilde{e}^u(x)$ and $x\tilde{e}^d(x)$ increase, indicating strong unpolarized-parton interferences at small $x$.
The genuine twist-3 $x$PDFs $x\tilde{g}^u_T(x)$ and $x\tilde{h}^u_L(x)$ exhibit nearly identical tendencies for $x\textgreater0.3$. The $d$ quark also displays a similar behavior.


In panel (b) of Fig.~\ref{fig:xpdf}, we observe that the twist-3 $x$PDFs have distinctly different $x$-dependence behaviors compared with the genuine twist-3 $x$PDFs. 
The amplitude of the twist-3 PDFs is larger than that of the genuine twist-3 PDFs, which indicates that the twist-2 PDFs strongly dominate the total twist-3 PDFs. 
Due to the $x$-dependent factor, $1/x$, in the EOM relation, such as that in Eq.~(\ref{eq:Eom1}), and the relatively small magnitude of genuine twist-3 $x$PDFs, the total twist-3 $x$PDFs exhibit a closer resemblance to the twist-2 PDFs.
Specifically, for small $x$, the twist-3 $x$PDFs $xe^u(x)$ and $xe^d(x)$ display a plateau, exhibiting a similar behavior to $f_1(x)$, \ie, the unpolarized PDFs $f_1^{u,d}(x)$ which exhibits a plateau at small $x$.
In this analysis, we refrain from comparing our result for the twist-3 PDF $e(x)$ with the experimental extraction~\cite{Courtoy:2022kca}. The reason lies in the fact that the extraction relies on the WW approximation, where the tilde term $\tilde{e}(x)$ is disregarded.

\begin{figure}[h]
  \centering 
  \includegraphics[width=0.38\textwidth]{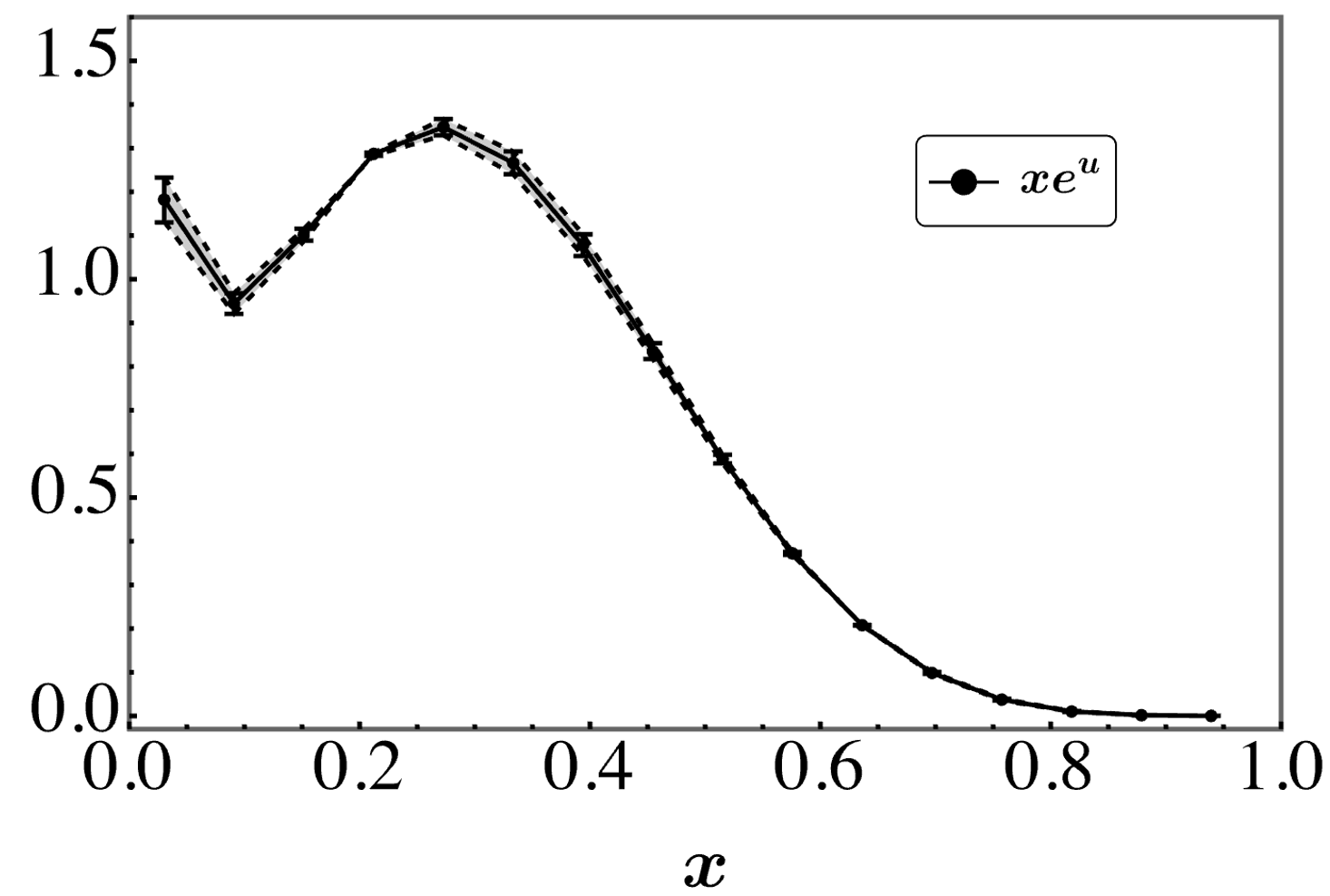} 
  \caption{Result for the unpolarized twist-3 $x$PDF, $xe^u(x)$. The uncertainties are quantified by comparing the results at different model masses of the $u$ quark.}
  \label{fig:epdf}
\end{figure}

In this work, we quantify the uncertainties by comparing the results at different model masses of the $u$ quark. 
More specifically, we shift the model mass of the $u$ quark from 0.31 GeV to 0.30 GeV and to 0.32 GeV, and then solve the light-front stationary Schr\"{o}dinger equation to obtain the two Fock-state LFWFs of the proton with the mass $M$ = 0.927 GeV and $M$ = 0.986 GeV, respectively. 
Finally, we evaluate the TMDs and PDFs at different $u$ quark masses to quantify the uncertainties. 
We present the result, including the uncertainties, for $xe^u(x)$ in Fig.~\ref{fig:epdf}. 
For the $xe^u(x)$, the largest relative uncertainty is 4.23\% at $x=0.5/16.5$. For the TMDs and remaining PDFs, the highest relative uncertainties do not surpass 5\%,
which demonstrates the stability of the model results with respect to slight changes in the input parameters.

It is important to note that the twist-3 TMDs and PDFs contain additional singular terms, $\delta(x)$, in the EOM relation when one considers the nontrivial gauge links and the zero modes~\cite{Efremov:2002qh}. These nontrivial contributions are vital for the sum rules,
\begin{align}
    \sum_q\int\mathrm{d}xe^q(x)&=\frac{\sigma_\pi}{m}\label{eq:sumrule1},\\
    \int\mathrm{d}xxe^q(x)&=\frac{m_q}{M}N_q,\label{eq:sumrule2}\\
    \int\mathrm{d}xg_1^q(x)&=\int\mathrm{d}xg_T^q(x)\label{eq:sumrule3},\\
    \int\mathrm{d}xh_1^q(x)&=\int\mathrm{d}xh_L^q(x)\label{eq:sumrule4},
\end{align}
where $m=\frac{1}{2}(m_u+m_d)$, and $\sigma_\pi$ is the pion-nucleon sigma term~\cite{Jaffe:1991ra,Efremov:2002qh}. 
The last two sum rules in Eqs.~(\ref{eq:sumrule3}) and (\ref{eq:sumrule4}) correspond to the well-known Burkhardt-Cottingham (BC) sum rules~\cite{Burkhardt:1970ti,Tangerman:1994bb,Burkardt:1995ts}. These two BC sum rules can be derived from the Lorentz invariant relations (LIRs), which are analyzed from the Lorentz structure of quark-quark correlations~\cite{Tangerman:1994bb}.

In this work, we simplify our study by neglecting zero modes. 
At the PDF level, the gauge link degenerates into a trivial identity matrix under the light cone gauge $A^+=0$~\cite{Belitsky:2002sm}, which has been considered here. 
Therefore, the only factor affecting the above sum rules, Eqs.~(\ref{eq:sumrule1}-\ref{eq:sumrule4}), is the contribution from the zero modes.
There is evidence that, with a one-loop dressed massive quark, $g_T(x)$ is the only exception and its BC sum rule given in Eq.~(\ref{eq:sumrule3}) is satisfied even when the singular term is not included~\cite{Burkardt:2001iy}. 
In contrast, both $e(x)$ and $h_L(x)$ fail to meet their respective sum rules when the origin is excluded from the integration region~\cite{Burkardt:2001iy}.
Table~\ref{tab:sumrule} shows our results of the quantities, including the uncertainties, on the left and right sides of the sum rules.
Although our results nearly satisfy the sum rule in Eq.~(\ref{eq:sumrule2}), a slight deviation is present, possibly attributed to the omission of the zero modes.
As mentioned in Ref.~\cite{Efremov:2002qh}, the pion-nucleon sigma term is completely related to singular terms or so-called zero mode contributions, which is proportional to $\delta(x)$. 
Therefore, in principle, the sigma term cannot be calculated without considering the zero modes.

Our results reveal significant deviations from the two BC sum rules, Eqs.~(\ref{eq:sumrule3}) and (\ref{eq:sumrule4}), indicating the crucial influence of singular terms for $g_T(x)$ and $h_L(x)$. 
The significance of this can be checked experimentally, as $g_T$ can be measured in the transversely polarized lepton-nucleon DIS~\cite{Kramer:2005qe,JeffersonLabHallA:2016neg,SANE:2018pwx} and $h_L$ appears in the longitudinal-transverse spin asymmetry in the polarized Drell-Yan process~\cite{Jaffe:1991kp,Koike:2008du} or single-inclusive particle production in nucleon-nucleon collisions~\cite{Koike:2016ura}.
An alternative explanation may lie in the truncation of the Fock space, leading to a violation of the Poincar\'e algebra~\cite{Mangin-Brinet:2003vmv}. 
This violation, in turn, disrupts the LIRs, ultimately causing the BC sum rules to be violated.


\begin{table}[h]
    \caption{BLFQ results of the quantities on the left and right sides of the sum rules given in Eqs.~(\ref{eq:sumrule2}-\ref{eq:sumrule4}). The uncertainties are quantified by comparing the results at different model masses of the $u$ quark.}
    \vspace{0.15cm}
    \label{para}
    \centering
    \begin{tabular}{ccc}
        \hline\hline
             & l.h.s. & r.h.s. \\        
        \hline 
        Eq.~(\ref{eq:sumrule2}) for $u$ quark & 0.647 $\pm$ 0.001 & 0.648 $\pm$ 0.001\\   
        Eq.~(\ref{eq:sumrule2}) for $d$ quark & 0.270 $\pm$ 0.008 & 0.261 $\pm$ 0.008\\   
        Eq.~(\ref{eq:sumrule3}) for $u$ quark & 0.796 $\pm$ 0.008 & 1.401 $\pm$ 0.029 \\
        Eq.~(\ref{eq:sumrule3}) for $d$ quark & -0.119 $\pm$ 0.007 & -0.412 $\pm$ 0.012 \\   
        Eq.~(\ref{eq:sumrule4}) for $u$ quark & 1.036 $\pm$ 0.008 & 0.035 $\pm$ 0.013\\  
        Eq.~(\ref{eq:sumrule4}) for $d$ quark & -0.314 $\pm$ 0.001 & 0.027 $\pm$ 0.002\\  
      \hline\hline
    \end{tabular}\label{tab:sumrule}
\end{table}

\section{Summary}
In this work, we conducted a systematic calculation of the subleading structure functions of the proton, specifically focusing on the interference terms, \ie, the genuine twist-3 distribution functions. 

We made approximations by neglecting zero modes and considering the gauge link as a unity matrix. These approximations result in the elimination of T-odd TMDs and the singular terms in the EOM relations.
First, we calculated the twist-2 T-even TMDs and the genuine twist-3 T-even TMDs, from the light-front wave functions (LFWFs) of the proton obtained within the Basis Light-Front Quantization (BLFQ) framework.
The genuine twist-3 distributions involve the overlaps of LFWFs of the $3q$ and $3q+g$ Fock components.
The LFWFs are obtained by solving the light-front QCD Hamiltonian in the light-cone gauge for the proton, truncated within the $|qqq\rangle$ and $|qqqg\rangle$ Fock spaces, and applying a confinement to the leading Fock sector. 
We then applied the EOM relations to obtain the total twist-3 T-even TMDs. Subsequently, by integrating the corresponding TMDs over $k_\perp$, we determined the twist-3 PDFs.

%
At the model scale, our results show that the genuine twist-3 distributions exhibit smaller values compared to the total twist-3 distributions, with noticeable values appearing at small $x$ especially $\tilde{e}$, which suggests the possibility of strong interference at lower longitudinal momentum fractions.
For the total twist-3 distributions, they are large at small $x$, implying that experiments at EicC with small $x$ and small transfer momentum $Q^2$ might observe high twist effects, such as $A_{LT}^{\cos\phi_S}$~\cite{Bacchetta:2006tn,Wang:2016dti}. 
An interesting aspect of our results is the evident violation of the BC sum rules, highlighting the significance of zero modes in the calculations of twist-3 distributions.
In addition, the satisfaction of the positivity bounds for twist-2 distributions and the independence of all the distributions suggest that the BLFQ framework is an effective tool for comprehending nonperturbative QCD.

We believe these results will further deepen our understanding of the partonic structures and the internal correlations of the proton, especially the interferences among multiple partons.
These results provide new nonperturbative input for future experiments to investigate the higher twist effects of the proton at the EIC and EicC.
In the future, we plan to include more Fock components, such as $3q+q\bar{q}$, and explore the sea quark distributions. 
Given the importance of the zero modes for the BC sum rules, we will also aim to investigate these aspects within the BLFQ framework.

\section*{Acknowledgements}

We thank Zuotang Liang and Tianbo Liu for valuable discussions.
Z. Zhu is supported by the Natural Science Foundation of Gansu Province, China, Grant No. 23JRRA571.
J. Lan is supported by the Special Research Assistant Funding Project, Chinese Academy of Sciences, by the National Science Foundation of Gansu Province, China, Grant No. 23JRRA631, and National Natural Science Foundation of China under Grant No. 12305095.
C. M. is supported by new faculty start up funding by the Institute of Modern Physics, Chinese Academy of Sciences, Grant No. E129952YR0.
X. Zhao is supported by new faculty startup funding by the Institute of Modern Physics, Chinese Academy of Sciences, by Key Research Program of Frontier Sciences, Chinese Academy of Sciences, Grant No. ZDBS-LY-7020, by the Natural Science Foundation of Gansu Province, China, Grant No. 20JR10RA067, by the Foundation for Key Talents of Gansu Province, by the Central Funds Guiding the Local Science and Technology Development of Gansu Province, Grant No. 22ZY1QA006, by international partnership program of the Chinese Academy of Sciences, Grant No. 016GJHZ2022103FN, by the Strategic Priority Research Program of the Chinese Academy of Sciences, Grant No. XDB34000000, and by the National Natural Science Foundation of China under Grant No.12375143.
J. P. V. is supported by the Department of Energy under Grant No. DE-SC0023692. 
This research is supported by Gansu International Collaboration and Talents Recruitment Base of Particle Physics (2023–2027).
A portion of the computational resources were also provided by Taiyuan Advanced Computing Center.

\appendix
\section{Overlap representations}
In this appendix, we provide an overview of the LFWF overlap representations for the twist-2 and twist-3 TMDs, and the genuine twist-3 TMDs of spin-$1/2$ baryons.

Through a rotation transformation~\cite{Lorce:2011zta}, we express the bound state of the baryon $|P, S\rangle$ in Eqs.~(\ref{eq:TMDs1}-\ref{eq:TMDs3}), (\ref{eq:e}-\ref{eq:h_L}), and (\ref{eq:gTMD1}-\ref{eq:gTMD3}) in terms of the light-front helicity state~\cite{Brodsky:1997de},
\begin{align}
    &|P,\Lambda\rangle=\sum_{\mathcal{N}}\sum_{\lambda_{1},\lambda_{2},\cdots,\lambda_{n}} \int \prod_{i}^{n}\frac{\left[\mathrm{d} x_{i} \mathrm{d}^2 \vec k_{\perp i}\right]}{2(2\pi)^3\sqrt{x_i}}\nonumber\\
    &\times2(2\pi)^3 \delta\left(1-\sum_{i}^{n} x_{i}\right)\delta^{2}\left(\vec P_{\perp}-\sum_{i}^{n} \vec k_{\perp i}\right) \nonumber\\
    &\times\Psi_{\mathcal{N},\{\lambda_{i}\}}^{\Lambda}(\{x_i, \vec k_{\perp i}\})\left|\{x_i P^+,\vec k_{\perp i} +x_i \vec P_{\perp},\lambda_i\}\right\rangle ,
    \label{BoundState0}
\end{align}
where $\Psi_{\mathcal{N},\{\lambda_{i}\}}^{\Lambda}$ is the LFWF in some specific Fock sector.
This rotation is represented as $(|P,+S\rangle, |P,-S\rangle)=(|P,+\rangle, |P,-\rangle)u(\theta,\varphi)$, where $u(\theta,\varphi)$ is the rotation transformation and the spin vector $S$ corresponds to the angle as $(S^1_\perp$, $S_\perp^2$, $S^3)$ = $(\sin\theta\cos\varphi$, $\sin\theta\sin\varphi$, $\cos\theta)$. 

Therefore, the correlators on the l.h.s. of Eqs. (\ref{eq:TMDs1}-\ref{eq:TMDs3}) and (\ref{eq:gTMD1}-\ref{eq:gTMD3}) can be expressed in terms of the light-front helicity amplitude as follows,
\begin{align}
    \Phi_{{\Lambda^\prime} \Lambda}^{[\Gamma]}&( x,  k_{\perp})=\left.\frac{1}{2} \int \frac{\mathrm{d} z^{-} \mathrm{d}^2 z^{\perp}}{2(2 \pi)^{3}} e^{i k \cdot z}\right.\nonumber\\
    &\times\left.\left\langle P, \Lambda^\prime\right|{\bar{\psi}}(0) \Gamma {\psi}(z)\left| P, \Lambda\right\rangle\right|_{z^{+}=0},\label{eq:TMDsdefinition}\\
    \Phi_{Ai;{\Lambda^\prime} \Lambda}^{[\Gamma]}&( x,  k_{\perp})=\left.\frac{1}{2}g_s \int \frac{\mathrm{d} z^{-} \mathrm{d}^2 z^{\perp}}{2(2 \pi)^{3}} e^{i k \cdot z}\right.\nonumber\\
    &\times\left.\left\langle P, \Lambda^\prime\right|{\bar{\psi}}(0) \Gamma A_{\perp i}(z){\psi}(z)\left| P, \Lambda\right\rangle\right|_{z^{+}=0}.\label{eq:TMDsdefinition2}
\end{align}

Using the rotation transformation along with a properly chosen polarization direction, the twist-2 TMDs and the genuine twist-3 TMDs can be expressed in terms of the light-front helicity amplitudes as,
\begin{align}
        f_1&=\frac{1}{2}\Big(\Phi_{++}^{[\gamma^+]}+\Phi_{--}^{[\gamma^+]}\Big)\label{eq:f_1},\\
        g_{1L}&=\frac{1}{2}\Big(\Phi_{++}^{[\gamma^+\gamma_5]}-\Phi_{--}^{[\gamma^+\gamma_5]}\Big),\\
        g_{1T}&=\frac{M}{k_\perp^1}\frac{1}{2}\Bigl(\Phi_{+-}^{[\gamma^+\gamma_5]}+\Phi_{-+}^{[\gamma^+\gamma_5]}\Bigr),\\
        h_1&=\frac{1}{4}\Big[\Bigl(\Phi_{+-}^{   [i\sigma^{1+}\gamma_5]}+\Phi_{-+}^{[i\sigma^{1+}\gamma_5]}\Bigr)\nonumber\\
        &+i\Bigl(\Phi_{+-}^{[i\sigma^{2+}\gamma_5]}-\Phi_{-+}^{[i\sigma^{2+}\gamma_5]}\Bigr)\Big],\\
        h_{1T}^\perp&=\frac{2M^2}{(k^1_\perp)^2-(k_\perp^2)^2}\frac{1}{4}\Big[\Bigl(\Phi_{+-}^{[i\sigma^{1+}\gamma_5]}+\Phi_{-+}^{[i\sigma^{1+}\gamma_5]}\Bigr)\nonumber\\
        &-i\Bigl(\Phi_{+-}^{[i\sigma^{2+}\gamma_5]}-\Phi_{-+}^{[i\sigma^{2+}\gamma_5]}\Bigr)\Big],\\
        h_{1L}^\perp&=\frac{M}{k^1_\perp}\frac{1}{2}\Bigl(\Phi_{++}^{[i\sigma^{1+}\gamma_5]}-\Phi_{--}^{[i\sigma^{1+}\gamma_5]}\Bigr),\\
        \tilde{e}&=\frac{1}{Mx}\Im m\Big(\tilde{\Phi}_{Ai,++}^{[\sigma^{i+}]}+\tilde{\Phi}_{Ai,--}^{[\sigma^{i+}]}\Big),\\
        \tilde{h}^\perp_T&=\frac{1}{2x(\vec k_\perp)^2}\Im m\Big(k_R\tilde{\Phi}_{Ai,+-}^{[\sigma^{i+}]}-k_L\tilde{\Phi}_{Ai,-+}^{[\sigma^{i+}]}\Big),\\
        \tilde{h}_{T}&=\frac{-1}{2x(\vec k_\perp)^2}\Im m\Big(k_R\tilde{\Phi}_{Ai,+-}^{[\sigma^{i+}\gamma_5]}+k_L\tilde{\Phi}_{Ai,-+}^{[\sigma^{i+}\gamma_5]}\Big),\\
        \tilde{h}_L&=-\frac{1}{2Mx}\Im m\Big(\tilde{\Phi}_{Ai,++}^{[\sigma^{i+}\gamma_5]}-\tilde{\Phi}_{Ai,--}^{[\sigma^{i+}\gamma_5]}\Big),\\
        \tilde{f}^\perp&=\frac{1}{2x(\vec k_\perp)^2}\nonumber\\
        \times&\Re e\Big\{k_{\perp}^1\big[\tilde{\Phi}_{Ai,++}^{[(-\delta^{1i}+i\epsilon_{\perp}^{1i}\gamma_5)\gamma^+]}+\tilde{\Phi}_{Ai,--}^{[(-\delta^{1i}+i\epsilon_{\perp}^{1i}\gamma_5)\gamma^+]}\big]\nonumber\\
        +&k_{\perp}^2\big[\tilde{\Phi}_{Ai,++}^{[(-\delta^{2i}+i\epsilon_{\perp}^{2i}\gamma_5)\gamma^+]}+\tilde{\Phi}_{Ai,--}^{[(-\delta^{2i}+i\epsilon_{\perp}^{2i}\gamma_5)\gamma^+]}\big]\Big\},\\
        \tilde{g}_L^\perp&=\frac{1}{2x(\vec k_\perp)^2}\nonumber\\
        \times&\Im m\Big\{k_{\perp}^1[\tilde{\Phi}_{Ai,++}^{[(-\delta^{2i}+i\epsilon^{2i}_\perp\gamma_5)\gamma^+]}-\tilde{\Phi}_{Ai,--}^{[(-\delta^{2i}+i\epsilon^{2i}_\perp\gamma_5)\gamma^+]}]\nonumber\\
        -&k_{\perp}^2[\tilde{\Phi}_{Ai,++}^{[(-\delta^{1i}+i\epsilon^{1i}_\perp\gamma_5)\gamma^+]}-\tilde{\Phi}_{Ai,--}^{[(-\delta^{1i}+i\epsilon^{1i}_\perp\gamma_5)\gamma^+]}]\Big\},\\
        {\tilde{g}_T}&=\frac{1}{4Mx}\nonumber\\
        \times&\Im m\Big\{i\tilde{\Phi}_{Ai,+-}^{[(-\delta^{1i}+i\epsilon_\perp^{1i}\gamma_5)\gamma^+]}-i\tilde{\Phi}_{Ai,-+}^{[(-\delta^{1i}+i\epsilon_\perp^{1i}\gamma_5)\gamma^+]}\nonumber\\
        &-\tilde{\Phi}_{Ai,+-}^{[(-\delta^{2i}+i\epsilon_\perp^{2i}\gamma_5)\gamma^+]}-\tilde{\Phi}_{Ai,-+}^{[(-\delta^{2i}+i\epsilon_\perp^{2i}\gamma_5)\gamma^+]}\Big\},\\
        \tilde{g}_T^\perp&=-\frac{M}{2x((k_{\perp}^1)^2-(k_{\perp }^2)^2)}\nonumber\\
        \times&\Im m\Big\{i\tilde{\Phi}_{Ai,+-}^{[(-\delta^{1i}+i\epsilon^{1i}\gamma_5)\gamma^+]}-i\tilde{\Phi}_{Ai,-+}^{[(-\delta^{1i}+i\epsilon^{1i}_\perp\gamma_5)\gamma^+]}\nonumber\\
        &+\tilde{\Phi}_{Ai,+-}^{[(-\delta^{2i}+i\epsilon^{2i}_\perp\gamma_5)\gamma^+]}+\tilde{\Phi}_{Ai,-+}^{[(-\delta^{2i}+i\epsilon^{2i}_\perp\gamma_5)\gamma^+]}\Big\}\label{eq:gtildeTperp},
\end{align}
where $k_{R,L}=k^1_\perp\pm ik^2_\perp$.

Substituting in the mode expansion of the quark and gluon field operators and the expression of the light-front helicity proton state, the light-front helicity amplitudes, given in Eqs.~(\ref{eq:TMDsdefinition}) and (\ref{eq:TMDsdefinition2}), can be expressed in terms of overlap of the LFWFs. We present these results                                                     as 
\begin{align}
    &\Phi^{[\Gamma]}_{\Lambda^\prime\Lambda}(x,k_\perp)=\sum_{{\lambda_1^\prime\lambda_1\lambda_2\lambda_3}}\int\mathrm{d}[123]\Psi^{\Lambda^\prime*}_{3,\lambda_1^\prime\lambda_2\lambda_3}\Psi^{\Lambda}_{3,\lambda_1\lambda_2\lambda_3}\nonumber\\
    &\times\frac{1}{2k_1^+}\bar{u}(p^\prime_1,\lambda_1^\prime)\Gamma u(p_1^\prime,\lambda_1)\delta^3(\tilde{k}-\tilde{p}_1^\prime)\nonumber\\
    &+\sum_{{\lambda_1^\prime\lambda_1\lambda_2\lambda_3\lambda_4}}\int\mathrm{d}[1234]\Psi^{\Lambda^\prime*}_{4,\lambda_1^\prime\lambda_2\lambda_3\lambda_4}\Psi^{\Lambda}_{4,\lambda_1\lambda_2\lambda_3\lambda_4}\nonumber\\
    &\times\frac{1}{2k_1^+}\bar{u}(p_1,\lambda_1^\prime)\Gamma u(p_1,\lambda_1)\delta^3(\tilde{k}-\tilde{p}_1),\label{eq:TMDoverlap}\\
    &\tilde{\Phi}_{Ai,\Lambda^\prime\Lambda}^{[\Gamma]}(x,k_\perp)=g_sC_F\sum_{{\lambda_1^\prime\lambda_1\lambda_2\lambda_3\lambda_4}}\int\mathrm{d}[123]\mathrm{d}[1234]\frac{[2(2\pi)^3]^2}{\sqrt{x_4}}\nonumber\\
    &\times\delta^3(\tilde{p}_2-\tilde{p}_2^\prime)\delta^3(\tilde{p}_3-\tilde{p}_3^\prime)\Big\{\Psi^{\Lambda^\prime*}_{3,\lambda_1^\prime\lambda_2\lambda_3}\Psi^{\Lambda}_{4,\lambda_1\lambda_2\lambda_3\lambda_4}\nonumber\\
    &\times\frac{1}{2k^+}\bar{u}(p_1^\prime,\lambda_1^\prime)\Gamma u(p_1,\lambda_1)\epsilon_{i}(\lambda_4)\delta^3(\tilde{k}-\tilde{p}_1^\prime)\nonumber\\
    &+\Psi^{\Lambda^\prime*}_{4,\lambda_1\lambda_2\lambda_3\lambda_4}\Psi^{\Lambda}_{3,\lambda_1^\prime\lambda_2\lambda_3}\nonumber\\
    &\times\frac{1}{2k^+}\bar{u}(p_1,\lambda_1)\Gamma u(p_1^\prime,\lambda^\prime_1)\epsilon^*_{i}(\lambda_4)\delta^3(\tilde{k}-\tilde{p}_1)\Big\},\label{eq:gTMDoverlap}
\end{align}
where we omit the arguments $(\tilde{p}_1^\prime,\tilde{p}_2^\prime,\tilde{p}_3^\prime )$ and $(\tilde{p}_1,\tilde{p}_2,\tilde{p}_3,\tilde{k}_4)$ from the LFWFs of the $|uud\rangle$ and $|uudg\rangle$ Fock sectors, respectively.
Here, $\tilde{p}_i\equiv(x_i,\vec{p}_{\perp i} )$ and $\tilde{p}^\prime_i\equiv(x^\prime_i,\vec{p}_{\perp i}^{\; \prime} )$. 
$\tilde{k}$ represents the arguments of TMDs.
$\tilde{p}_1^\prime$ and $\tilde{p}_1$ denote the momenta of the struck quark in the $|uud\rangle$ and $|uudg\rangle$ Fock sectors, respectively.
$\tilde{p}_2$, $\tilde{p}_3$, $\tilde{p}_2^\prime$ and $\tilde{p}_3^\prime$ represent the momenta of the spectator quarks.
$\tilde{p}_4$ represents the momentum of the gluon.
$\lambda_1^\prime$ and $\lambda_1$ represent the helicities of the struck quark in the $|uud\rangle$ and $|uudg\rangle$ Fock sectors, respectively.
$\lambda_2$ and $\lambda_3$ represent the helicities of the spectator quarks.
$\lambda_4$ represents the helicity of the gluon.
$\bar{u}(k,\lambda)$ and $u(k,\lambda)$ represent the spinors.
$C_F$ is the color factor.
$\epsilon_i(\lambda)$ is the polarization vector.
We define the integral symbol as, 
\begin{align}
    \mathrm{d}[123]\equiv \frac{\prod_1^3\mathrm{d}x_i^\prime\mathrm{d}^2\vec{p}_{\perp i}^{\;\prime}}{[2(2\pi)^3]^3}2(2\pi)^3\delta^3(\tilde{P}-\sum_{i=1}^3\tilde{p}_i^\prime),\\
    \mathrm{d}[1234]\equiv \frac{\prod_1^4\mathrm{d}x_i\mathrm{d}^2\vec{p}_{\perp i}^{}}{[2(2\pi)^3]^4}2(2\pi)^3\delta^3(\tilde{P}-\sum_{i=1}^4\tilde{p}_i),
\end{align}
where $\delta^3(\tilde{P}-\sum_{1}^n\tilde{p}_i)=\delta(1-\sum_1^n x_i)\delta^2(\vec {P}_\perp-\sum_1^n\vec{p}_{\perp i})$. 

The above derivation is based on the unit approximations for the gauge links, $\mathcal{U}^{n_-}_{(0,+\infty)}\approx \mathbbm{1}$ and $\mathcal{U}^{n_-}_{(+\infty,z)}\approx \mathbbm{1}$. In this work, we do not take into account the non-trivial contribution of the gauge link, which would introduce an additional phase, leading to the presence of non-zero T-odd TMDs~\cite{Belitsky:2002sm}.

\biboptions{sort&compress}
\bibliographystyle{elsarticle-num}
\bibliography{TMDsRef.bib}
\end{document}